\documentclass[twocolumn,superscriptaddress,super,floatfix]{revtex4-2}
\usepackage[utf8]{inputenc}
\usepackage[english]{babel}
\usepackage{amsmath}
\usepackage{braket}
\usepackage{bm}
\usepackage{chemformula}
\usepackage{siunitx}
\usepackage{amsbsy}
\usepackage{float}
\usepackage[version=4]{mhchem}

\usepackage{graphicx}
\usepackage{amssymb}
\usepackage{amsfonts,dsfont}
\usepackage{subfigure}

\usepackage[normalem]{ulem}
\usepackage[bbgreekl]{mathbbol}
\usepackage{mathrsfs,empheq}
\usepackage{relsize,scalerel}

\usepackage[toc,page]{appendix}
\usepackage{hyperref}
\usepackage[savepos]{zref}

\usepackage{siunitx}[=v2]

\begin{document}

\title{Quantum effects in the H-bond symmetrization and in the thermodynamic properties of high pressure ice}

\author{Marco Cherubini}
\email{marco.cherubini@sorbonne-universite.fr}
\affiliation{Institut de Minéralogie, de Physique de Matériaux et de Cosmochimie, CNRS, Sorbonne Université, 4 Place Jussieu,Paris,75005,France}

\author{Lorenzo Monacelli}
\affiliation{Theory and Simulation of Materials (THEOS),
Ecole Polytechnique Fédérale de Lausanne, CH-1015 Lausanne, Switzerland}
\affiliation{Dipartimento di Fisica, Università di Roma Sapienza, Piazzale Aldo Moro 5, I-00185
Roma, Italy}

\author{Bingjia Yang}
\affiliation{Department of Chemistry, Princeton University, Princeton, New Jersey 08544, USA}

\author{Roberto Car}
\affiliation{Department of Chemistry, Princeton University, Princeton, New Jersey 08544, USA}

\author{Michele Casula}
\email{michele.casula@sorbonne-universite.fr}
\affiliation{Institut de Minéralogie, de Physique de Matériaux et de Cosmochimie, CNRS, Sorbonne Université, 4 Place Jussieu,Paris,75005,France}

\author{Francesco Mauri}
\email{francesco.mauri@uniroma1.it}
\affiliation{Dipartimento di Fisica, Università di Roma Sapienza, Piazzale Aldo Moro 5, I-00185
Roma, Italy}
\date{\today}

\begin{abstract}
We investigate the structural and thermodynamic properties of high-pressure ice by incorporating quantum anharmonicity at a non-perturbative level. Quantum fluctuations reduce the critical pressure of the phase transition between phase VIII (with asymmetric H-bonds) and phase X (with symmetric H-bonds) by 65 GPa from its classical value of 116 GPa at 0K. Moreover, quantum effects make it temperature-independent over a wide temperature range (0K-300K), in agreement with experimental estimates obtained through vibrational spectroscopy and in striking contrast to the strong temperature dependence found in the classical approximation. The equation of state shows fingerprints of the transition in accordance with experimental evidence. Additionally, we demonstrate that, within our approach, proton disorder in phase VII has a negligible impact on the occurrence of phase X. Finally, we reproduce with high accuracy the 10 GPa isotope shift due to the hydrogen-to-deuterium substitution. 
\end{abstract}

\maketitle

\section{Introduction}
\label{sec:Introduction}

Water ice is a unique material with one of the richest phase diagrams existing in nature \cite{Leadbetter1985,Kamb1964,Kamb1968,Engelhardt1981,Kamb1967,DOWELL1960,Pruzan1993}. In stable structures at low pressures, every water molecule forms hydrogen (H) bonds with its four nearest neighbors. In denser phases such as VII, VIII, and X \cite{Pruzan1993,Benoit1998,Pruzan2003}, two interpenetrated but not interconnected hydrogen bond networks are present. Both ice VIII and its proton disordered counterpart ice VII, are expected to exhibit proton symmetrization under pressure leading to phase X\cite{Pruzan1993}. Nevertheless, the experimental evidence of proton symmetrization is only indirect due to the difficulties in accurately detecting proton positions at high pressure. Theory-wise, a precise structural description of ice phases is challenging, as nuclear quantum effects must be properly included. Consequently, the occurrence of hydrogen bond symmetrization in ice still lacks a definitive picture despite years of theoretical and experimental research. \cite{Hirsch1984,Hemley1987,Benoit1998}. The focus of the present work is to shed light on this phenomenon within the framework of the stochastic self-consistent harmonic approximation (SSCHA) \cite{Monacelli2021}, ideally suited to describe transitions driven by strong quantum anharmonicity.

\begin{figure*}
    \centering
        \includegraphics[width=\textwidth]{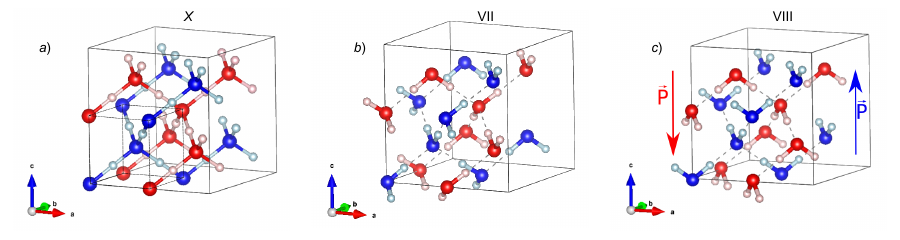}
    \caption{High-pressure dense phases of ice. The two interpenetrated but not interconnected hydrogen bond networks are shown by different colors in the figure (hydrogen and oxygen atoms represented by small and large balls, respectively). Within each network, water molecules are linked in tetrahedral coordination through weak hydrogen bonds (depicted by dashed lines). All structures are presented with an identical number of atoms (16 water molecules; 8 per network), corresponding to those employed in our numerical simulations. Panel a): 2x2x2  supercell of ice X. The bcc unit cell, containing only two water molecules, is shown by the dashed black lines. Panel b): Bcc unit cell of ice VII. In this disordered proton configuration, hydrogen atoms are randomly oriented, fulfilling the ice rules \cite{Bernal1933}, and no net polarization is present. Panel c): Supercell of tetragonal ice VIII. Protons in the two hydrogen bond networks align along the vertical c-axis, forming a macroscopic net polarization $\vec{P}$ (arrows whose color is associated with the corresponding network) with opposite directions to generate the antiferroelectric structure}
    \label{fig:structures}
\end{figure*}

The H-bond symmetrization emerges under pressure from the correlation between the covalent bond length $d_{\ce{OH}}$ and the inter-molecular distance $d_{\ce{OO}}$. Under compression, a nonlinear covalent bond stretching is observed as $d_{\ce{OO}}$ decreases \cite{Novak,Ichikawa1978} until the proton occupies the midpoint between the two nearest neighbor oxygen atoms \cite{Zhang2016}. These structural properties are supported by the observation of the softening of the stretching mode frequencies \cite{Goncharov1996,Aoki1996,Goncharov1999}. In the symmetric phase, referred to as ice X, the arrangement of ice is isomorphic to the cuprite ($\ch{Cu_2O}$) structure, and the conventional picture of ice as a molecular crystal breaks down. Ice X adopts a body-centered cubic (bcc) structure with the space group $Pn\bar{3}m$, and its unit cell accommodates two water molecules in high symmetry positions. The $2\times2\times2$ supercell of ice X used in this work is illustrated in Fig. \ref{fig:structures} (a), where the two interpenetrated sublattices are highlighted by different colors.

In the low-pressure region of the phase diagram, preceding the formation of ice X, the predominant structures are phases VII and VIII, related to each other by an order-disorder transition. Ice VII exhibits orientational proton disorder and has a cubic paraelectric structure \cite{Bridgman1937}. In the disordered configurations, various proton arrangements satisfy the Bernal-Fowler ice rules \cite{Bernal1933}. In Fig. \ref{fig:structures} (b), we depict one such configuration used in our simulations. It has been observed that ice VII orders into phase VIII upon cooling \cite{Whalley1966}. The lattice of the ordered structure is obtained by slightly distorting the cubic phase VII into a tetragonal unit cell, followed by a displacement of one of the two sublattices in the $\vec{c}$ axis direction (see Fig. \ref{fig:structures} (c)). In this structure, hydrogen atoms align along a preferential direction to generate a macroscopic net polarization $\vec{P}$ with opposite signs in the two sublattices, resulting in an antiferroelectric arrangement.

\begin{figure}
    \centering
    \includegraphics[width=\columnwidth]{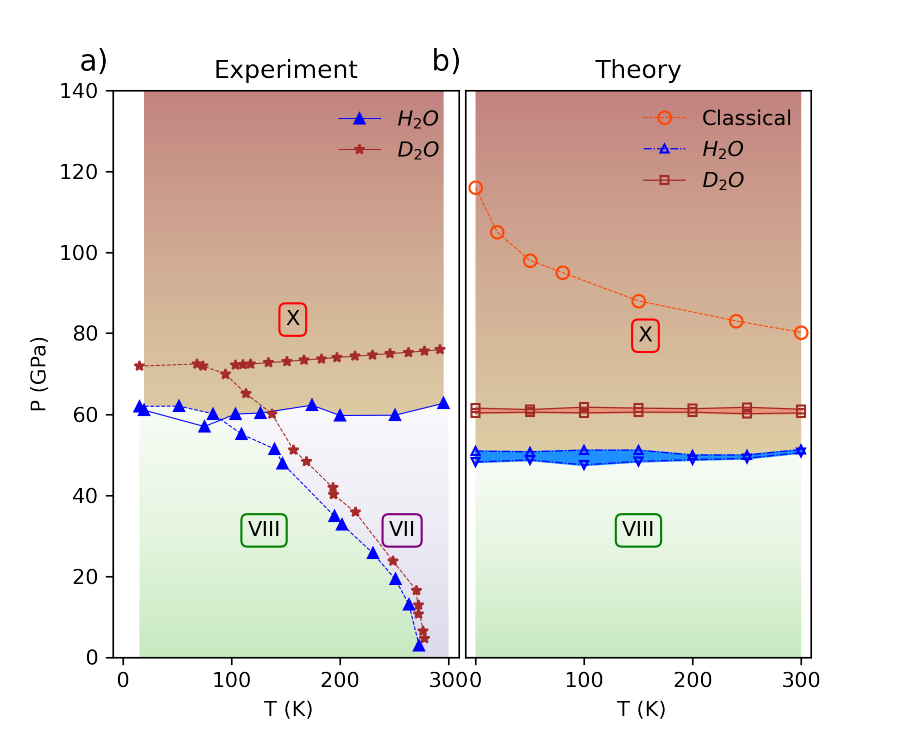}
    \caption{Panel a): Experimental phase diagram for high-pressure ice.  The horizontal upper blue triangles and the brown stars indicate the H-bond symmetrization transition for $\ce{H_2O}$ and $\ce{D_2O}$ ice, respectively, determined through Raman spectroscopy \cite{Goncharov1999}. Same symbols, but connected 
    by dashed lines, indicate the VII-VIII phase boundary \cite{Pruzan2003} for the two isotopes. The experimental stability domain of each phase, for $\ce{H_2O}$ ices, is shown in the figure with shaded areas of different colors. Panel b): Theoretical results obtained by SSCHA simulations for the VIII-X transition. Blue and brown narrow bands show the computed critical pressure for $\ce{H_2O}$ and $\ce{D_2O}$ ices, respectively. The band boundaries correspond to the extrema of the hysteresis cycle. The stability domains of ice X and ice VIII at low pressure, for $\ce{H_2O}$ ice, are shaded in pink and green, respectively. The classical VIII-X transition is included for comparison (red empty circles).}
    \label{fig:final_plot}
\end{figure}

Over the last few decades, the VII-VIII phase boundary has been extensively studied and is now rather well-established \cite{Pruzan1993,Klotz2017,Yoshimura2006,Komatsu_2020,Guthrie2013}. The temperature dependence of the critical pressure exhibits distinct regimes (see Fig. \ref{fig:final_plot} (a)). At high temperatures, the transition is predominantly influenced by thermal effects, where entropy favors the stability of the disordered phase VII. Conversely, as the system is cooled, the VII-VIII transition is fully governed by quantum effects\cite{Pruzan2003}. 

The underlying proton potential energy surface (PES) exhibits a double-well shape at low pressure, gradually transforming upon compression towards a single-well potential associated with the symmetric phase X \cite{Pruzan1993,Benoit1998,Benoit1998_2}. However, the phase boundary between ice VII and ice X is affected by significant uncertainties, with reported critical pressures ranging from 30 GPa to 120 GPa \cite{Grande2022,Wolanin1997,Hama1994,Loubeyre1999,Pruzan1993,Sugimura2008,Mndez2021,Asahara2010,Guthrie2019,Meier2018,Fukui2022,Zha2016,Goncharov1996,Goncharov1999,Song1999,Song2003,Struzhkin1997,Pruzan2003,Aoki1996,Hirsch1984}.
The fact that phases VII and X share the same symmetry group makes difficult to identify experimental observables which allows to determine unambiguously the phase transition. As a result, different techniques lead to different outcomes, resulting in a broad range of estimations for the transition pressure. The existence of a gradual nature of the transition, which could allow for numerous intermediate structures in the path towards hydrogen bond symmetrization, has been hypothesized \cite{Wolanin1997,Sugimura2008,Meier2018,Mndez2021}.

Various structural measurements, including Brillouin scattering, x-ray, and neutron diffraction, aim at tracking the ice X transition by monitoring the disappearance of specific diffraction lines associated with low-pressure phases. Room temperature (RT) experiments reveal variations in hydrogen bond states at pressures ranging between 40 and 60 GPa \cite{Asahara2010,Loubeyre1999,Fukui2022}, indicative of hydrogen bond symmetrization. Nevertheless, other changes observed at lower pressures were considered to be related to intermediate states \cite{Asahara2010}. An upper pressure limit of 110 GPa for ice X formation has been derived from extrapolating neutron diffraction in phase VII \cite{Guthrie2019}. Furthermore, hints at the phase transition can be gleaned by examining changes in the equation of state (EOS), which is directly accessible through diffraction experiments. At RT, small discontinuities are observed in the region above 60 GPa \cite{Wolanin1997,Hama1994,Sugimura2008}, where the EOS closely follows the theoretically simulated one for ice X \cite{Sugimura2008,Asahara2010}. Conversely, at pressures lower than 40 GPa, the experimental EOS of phase VII agrees with the theoretical predictions for the same phase. For pressures in between 40 GPa and 60 GPa, observed EOS variations have been attributed to the presence of intermediate states \cite{Sugimura2008}.

The hydrogen bond transformation is also reflected in elastic properties, resulting in a lower compressibility for ice X. Measured bulk modulus values \cite{Mndez2021} exhibit different regimes as a function of pressure, associated with transitions to intermediate states (at 35 GPa and 50 GPa) and ultimately to symmetric ice X for $P\simeq$ 90-100 GPa. Moreover, proton nuclear magnetic resonance experiments \cite{Meier2018} provided a detailed model of the energy potential and detected a transition at about 75 GPa, consistent with ice X formation.

Finally, strong signatures of proton symmetrization are observable through both Raman \cite{Goncharov1999,Pruzan1993,Pruzan2003,Zha2016} and infrared (IR)\cite{Goncharov1996,Aoki1996,Song1999,Song2003,Struzhkin1997} spectroscopy. According to group theory, the phonon activity undergoes drastic changes across the transition. In contrast to the multitude of active modes in low-pressure phases, only two IR modes and a single Raman mode are active in ice X. Major transformations in the spectra, associated with features related to the cuprite structure, have been detected between 60 and 70 GPa \cite{Aoki1996,Struzhkin1997,Goncharov1996,Goncharov1999,Song1999,Song2003} (Fig. \ref{fig:final_plot} (a) shows the experimental symmetrization pressure obtained through Raman spectroscopy \cite{Goncharov1999}). Moreover, experimental measurements also indicate that temperature plays a negligible role. Indeed, the observed variations in the spectra occur at approximately the same pressures over a large temperature range spanning the stability domain of phases VII and VIII \cite{Pruzan1993,Goncharov1996,Song2003}. Although the pressure evolution of the vibrational peaks can be interpreted based on a proton symmetrization mechanism, their broadening and asymmetric shape across the transition are suggestive of the presence of intermediate states \cite{Song2003,Goncharov1996,Song1999,Zha2016,Benoit2002}, without being conclusive \cite{Aoki1996}.

It is well-known that nuclear quantum effects (NQE) play a pivotal role in the hydrogen bond symmetrization process of high-pressure ice, as evidenced by the 10 GPa increase in critical pressure upon isotopic substitution \cite{Wolanin1997,Pruzan2003,Song1999}. Quantum mechanical simulations demonstrate that the inclusion of NQE is essential for an accurate description of the phase transition, lowering the classical critical pressure \cite{Benoit1998,Bronstein2014,Benoit2002}. In path integral molecular dynamics (PIMD) simulations at $\approx$ 100K \cite{Benoit1998,Benoit1998_2}, the symmetrization is detected through the analysis of the proton distribution along the O-O direction. At RT, quantum thermal bath simulations \cite{Bronstein2014}, despite the approximations involved, reproduce typical spectroscopic features of ice X at around 60 GPa, even though the underlying PES is still of a double-well type. An extended quantum description of the hydrogen bond symmetrization covering a dense grid of temperatures and pressures is still lacking. This is necessary to elucidate the dynamics and nature of the transition. In this study, we account for both thermal and quantum nuclear fluctuations through numerical simulations within the SSCHA \cite{Monacelli2021}, spanning a broad temperature range (0K-300K). The SSCHA method, grounded in a quantum variational minimization of the Gibbs free energy, assumes that the quantum wave functions of the nuclei can be effectively represented as a multidimensional Gaussian. Although this approach involves approximations compared to more exact theories like PIMD, it significantly reduces computational demands. Moreover, the SSCHA allows one to constrain the symmetry group of the relaxed structures granting a clear identification of the ice phases also close to the transition. Additionally, it enables us to track atomic positions systematically with changing pressure, providing a direct means of detecting proton symmetrizationthrough the evolution of the Gaussian centers. Finally, the vibrational free energy, accessible within the SSCHA formalism, allows one to compute the EOS across the transition with full inclusion of NQE.

The paper is organized as follows. In Section \ref{sec:methods}, we introduce the theoretical framework employed for the numerical simulations. We present our results valid for classical nuclei in Section \ref{sec:classical_results}, while we focus on the quantum regime and the effect of proton disorder in Section \ref{sec:quantum_results}. Conclusions are drawn in Section \ref{sec:conclusion}. Additional information on convergence tests and computational details is provided Appendices $\ref{app:convergence}$ and $\ref{app:hessian}$.

\section{Computational Methods}
\label{sec:methods}

We work within the Born-Oppenheimer approximation \cite{Born1927} to separate electronic and nuclear degrees of freedom. The total electronic energy, at fixed nuclei, is computed using a
Neural Network Potential (NNP) trained on the Perdew-Burke-Ernzerhof (PBE) \cite{Perdew1996} parametrization of the Density Functional Theory (DFT) exchange-correlation functional. A detailed description of the construction and validation of the NNP adopted can be found in App. \ref{app:DP}.
The nuclear quantum Hamiltonian is solved using the SSCHA theory \cite{Monacelli2021} which grants the inclusion of quantum and thermal fluctuations variationally. Our convergence tests, detailed in Appendix \ref{app:convergence}, prove that unit cells comprising 16 water molecules are adequate to attain the required convergence. We fix the temperature and carry out two distinct sets of simulations, each one exploring independently the ordered VIII and disordered VII ices. We iteratively relax the structures at increasing target pressures until the two low-pressure phases transform into ice X. When computing the presence of hysteresis effects, we reverse the procedure by starting from the high-pressure phase X and gradually decompressing the system with small pressure steps until the corresponding lower pressure structures are recovered. To accelerate the SSCHA minimization convergence, and to detect via the hysteresis cycle the presence of barriers between phases, at each pressure, we use previously relaxed geometries for the closest pressure value (from the above step in decompression and from below in compression) as a starting guess. Equivalently, the closure point of the hysteresis cycle can be spotted by looking at the disappearing of the local minimum in the free energy profile associated with the symmetric position. This feature can be pinpointed by computing the SSCHA free energy curvature in ice X at the O-O midpoint as a function of pressure. In its computation, we included both the third- and fourth-order force constant matrix in the self-energy correction in the static approximation (see Ref. \cite{Monacelli2021} for further details). 
This allows the calculation of the so-called full free energy Hessian that will exhibit imaginary frequencies when the aforementioned local minimum disappears and the free energy acquires a double-well shape. Therefore, a simple linear fit of the soft modes in the vicinity of the critical pressure allows us to determine the closure point of the hysteresis cycle. This procedure is computationally efficient since it requires the calculation of the Hessian in the high-symmetry phase X. Thus, after checking that these two methods yield consistent results (see Fig. \ref{appfig:hessian_hysteresis} in App. \ref{app:hessian}), we opt for using the free energy curvature. Conversely, to identify the starting point of the hysteresis cycle, we look at the centroid positions evolution upon compression. 

Concerning the disordered phase VII, we generate five distinct samples from the 626 possible hydrogen arrangements for a 16-water molecules structure \cite{Nagle1966}, employing the GenIce package \cite{Matsumoto2017}. After confirming the similarity of their outcomes in terms of symmetrization pressure (see Fig. \ref{fig:different_VIIs}), we present the results for only one of them, dubbed \emph{struct 0} as guideline throughout this study. For the SSCHA simulations, we employed a database of 7000 configurations to minimize and relax the structures at the target pressure with the necessary accuracy. Instead, to compute the full free energy Hessian, 50000 configurations were required to achieve convergence.

\section{Results}
\label{sec:results}

\subsection{Hydrogen bond symmetrization in the classical approximation}
\label{sec:classical_results}

The shortcomings of the classical approximation in capturing hydrogen bond symmetrization in water ice are well known. Previous studies have demonstrated that the incorporation of NQE significantly decreases the critical pressure \cite{Benoit1998, Bronstein2014, Benoit2002}. Nevertheless, as of now, there is a lack of investigations into how this effect depends on temperature. We explored the pressure evolution of the ordered ice VIII structure at various temperatures within the classical regime using the SSCHA theory. To achieve this, we increased the atomic masses by six orders of magnitude to eliminate quantum nuclear fluctuations while retaining thermal ones. 

\begin{figure}
    \centering
    \includegraphics[width=\columnwidth]{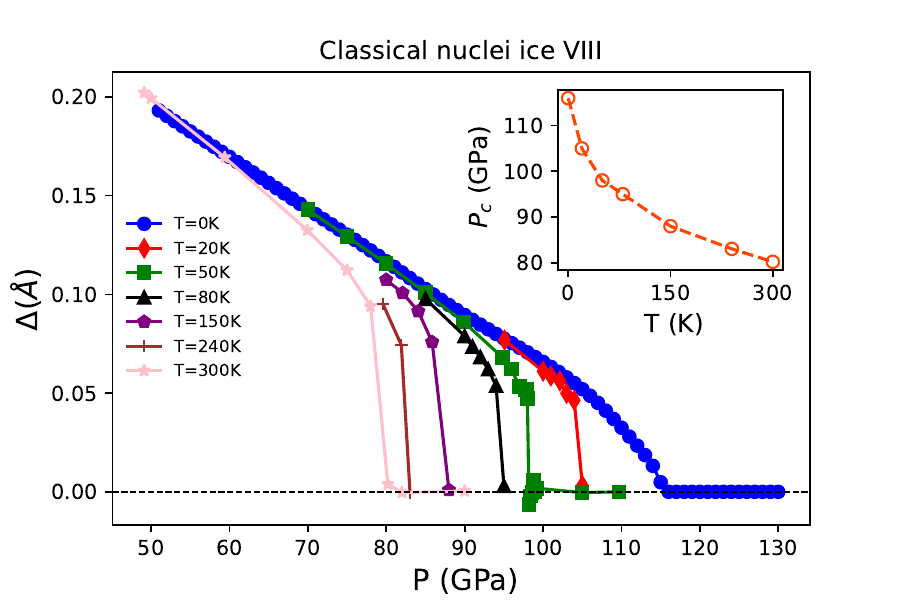}
    \caption{Order parameter in ice VIII in the classical approximation as function of pressure and temperature. In the main panel, a spectrum of colors and symbols, ranging from blue circles to pink stars, illustrates the order parameter $\Delta = |d_{\ce{OH}} - d_{\ce{OO}}/2| $ as a function of pressure and temperature, spanning from zero to room temperature. The horizontal grey dashed line serves as a reference for the zero value, corresponding to symmetric ice X. In the inset, the relative critical pressure, extracted from the order parameter, is pictured as a function of temperature by red empty circles. Error bars are too small ($\simeq$ 0.5 GPa) to be seen on the figure's scale.}
    \label{fig:Classical}
\end{figure}

Within the SSCHA, the possibility to directly monitor the centroid position throughout the minimization process as a function of pressure offers a convenient order parameter for characterizing transitions driven by proton symmetrization. We define our order parameter as $\Delta = | d_{\ce{OH}} - \frac{1}{2}d_{\ce{OO}} |$, representing the distance of the proton from the O-O midpoint, a quantity that vanishes in the symmetric phase X while it is non-zero in the symmetry broken phases VII and VIII. In Fig. \ref{fig:Classical}, we plot its variation with pressure in ice VIII for several temperatures, ranging from T=0K to T=300K. The inset of the same Figure shows the resulting temperature dependence of the classical critical pressure. In contrast to the quantum case (see Sec. \ref{sec:quantum_results}) and experimental observations \cite{Pruzan1993,Goncharov1996}, where the critical pressure exhibits negligible temperature dependence, the classical regime displays a pronounced nonlinear behavior. In the classical approximation, thermal nuclear fluctuations are so strong to induce a $\simeq$ 30$\%$ reduction of the critical pressure from T=0K to RT. The most substantial effect is observed 0K-50K range, where the transition pressure decreases drastically, from $P_c$(0K) = 116 GPa to $P_c$(50K) = 98 GPa, constituting 50 $\%$ of the total variation within the investigated temperature range. As the system is further heated, above 150 K, the temperature dependence becomes smoother, and only an 8 GPa reduction is observed, reaching $P_c=80$ GPa at T=300K.

To characterize the order of the phase transition at the classical level, we checked whether hysteresis effects are present in the VIII-to-X transformation, adopting the procedure outlined in Sec. \ref{sec:methods}. We performed this analysis at two temperatures, T=20K and T=300K, revealing the presence of a hysteresis cycle, whose pressure width diminishes upon heating, decreasing from $\simeq 3$ GPa at 20K to $\simeq$ 1 GPa at RT. Thus, the classical phase transition between the phase VIII and X is of a weakly first-order type.

\subsection{The effect of quantum nuclear fluctuations and proton disorder}
\label{sec:quantum_results}

\begin{figure}[ht!]
    \centering
    \includegraphics[width=0.9\columnwidth]{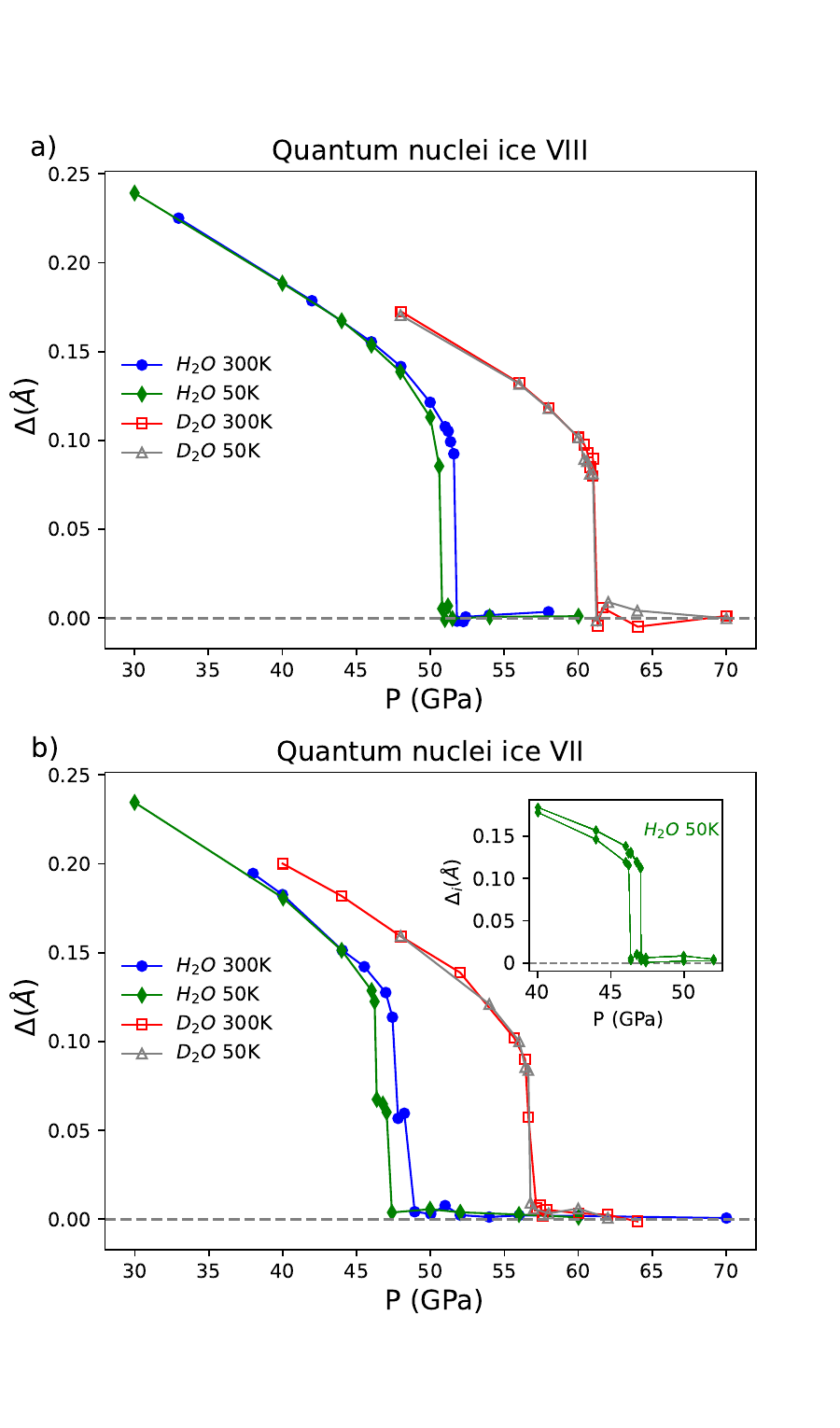}
    \caption{ Panel a): Order parameter for phase VIII in $\ce{H_2O}$ at 50K (filled green diamonds) and at 300K (filled blue circles), and in $\ce{D_2O}$ at 50K (empty grey upper triangles) and at 300K (empty red squares). Panel b): The same as panel a) but for ice VII. The inset in panel b) provides details on the local order parameter $\Delta_i$ computed for every hydrogen atom in the disordered unit cell at 50K. Two sets of values, slightly differing from each other, can be distinguished from our structures.}
    
    \label{fig:quantum_SSCHA}
\end{figure}

The classical picture is inadequate for accurately reproducing the experimental measurements. To account for NQE, we follow the same procedure adopted for the classical approximation but with the physical atomic masses. As outlined in Sec. \ref{sec:methods}, we carry out two independent sets of simulations for the ordered and disordered phases. In each set, we fix the temperature and progressively increase the pressure, relaxing the structures until achieving symmetrization of all hydrogen bonds in the simulation cell. We use the same order parameter $\Delta$ as for the classical regime of Fig. \ref{fig:Classical}. Its evolution with pressure and temperature is plotted in Fig. \ref{fig:quantum_SSCHA} for both phase VII (Fig. \ref{fig:quantum_SSCHA} (a)) and VIII (Fig. \ref{fig:quantum_SSCHA} (b)). Results for two temperatures, T=50K and T=300K, are presented.

First, we notice that temperature plays a marginal role, consistent with findings from various experiments \cite{Pruzan1993,Goncharov1996}. The discrepancy in critical pressures between T=50K and RT is approximately 1 GPa, in stark contrast to the pronounced nonlinearity observed in the classical approximation (see inset of Fig. \ref{fig:Classical}). This outcome underscores the strength of NQE in water ice, which dominate over thermal fluctuations. To give further reliability to this claim, we analyzed the VIII-X transition with a finer temperature grid, ranging from T=0K to RT by steps of 50K each (see Fig. \ref{fig:final_plot} (b)). This investigation confirms the temperature independence of the quantum hydrogen bond symmetrization pressure. Notably, we identified an approximately 60 GPa reduction in the quantum critical pressure compared to the classical one at T=0K, consistent with other theoretical predictions \cite{Benoit1998}.

Moreover, we assessed the impact of isotopic substitution by replicating the procedure with heavy water ice. In the case of $\ce{D_2O}$ ice, we observe a similar temperature independence of the VIII-X boundaries. The estimated critical pressure for $\ce{D_2O}$ ice is approximately 10 GPa higher than that of $\ce{H_2O}$ ice, in agreement with experimental data \cite{Wolanin1997,Pruzan2003,Song1999}.

At very low temperatures, simulations focusing only on ice VIII are sufficient to comprehensively characterize the hydrogen bond symmetrization transition. Conversely, as temperature increases, entropic effects come into play, favouring the appearance of the disordered phase. Therefore, having an accurate description of proton disorder is of paramount importance to investigate the transition towards the symmetric phase X at higher temperatures. In Fig. \ref{fig:quantum_SSCHA} (b), we present results for one of the disordered configurations fulfilling the ice rules, denoted hereafter as \emph{struct 0}. Unlike ice VIII, where proton order ensures uniform covalent bond lengths for all hydrogen atoms, the disordered phase requires the definition of a local order parameter $\Delta_i$ for each hydrogen atom $i$ within the structure. Ultimately, we establish the global order parameter $\Delta$ for ice VII by averaging over the local order parameters. $\Delta$ is plotted as a function of pressure in Fig. \ref{fig:quantum_SSCHA} (b). Similarly to ice VIII, temperature plays a minor role in the hydrogen bond symmetrization in ice VII, contributing solely to a 1 GPa difference in the critical pressure between T=50K and RT. Moreover, the size of the isotopic effect is consistent with that observed in the ordered ice and with experimental findings. Indeed, the phase VII order parameter is characterized by a step-like behavior in approaching the transition due to the non-homogeneous distribution of the hydrogen bonds within the disordered structure. The steps can be attributed to a situation where only a subset of the full supercell underwent a hydrogen bond symmetrization, while other bonds contribute to a non-vanishing order parameter $\Delta$. The non-homogeneous evolution of the local order parameters $\Delta_i$ as a function of pressure is shown in the inset of Fig. \ref{fig:quantum_SSCHA} (b). 

\begin{figure}
    \centering
    \includegraphics[width=0.45\textwidth]{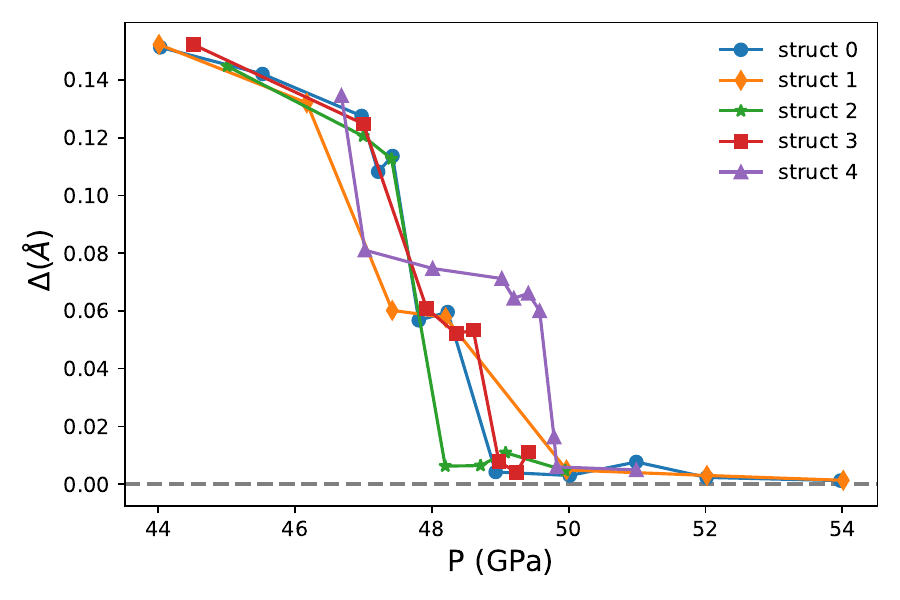}
    \caption{Order parameter for 5 different proton disorder configurations of the ice VII structure at 300K. The grey dashed horizontal line corresponds to the symmetric ice X. }
    \label{fig:different_VIIs}
\end{figure}

To examine the dependence of the symmetrization pressure on various realizations of proton disorder in ice VII, we analyzed five different hydrogen arrangements. In Fig. \ref{fig:different_VIIs}, we present their global order parameters as a function of pressure. Based on this statistics, we observe that the occurrence of this step-like behavior is restricted in a narrow pressure range of 2 GPa
below the global phase transition. Within our SSCHA approach, we can thus argue that the hypothesized intermediate phases mentioned in the Introduction cannot be identified with this phenomenon.  Indeed, experimental evidence suggests that if such intermediate states exist, they should appear in a pressure window of the order of 10-20 GPa, significantly larger than the range observed in our simulations.

\begin{figure}
    \centering
    \includegraphics[width=\columnwidth]{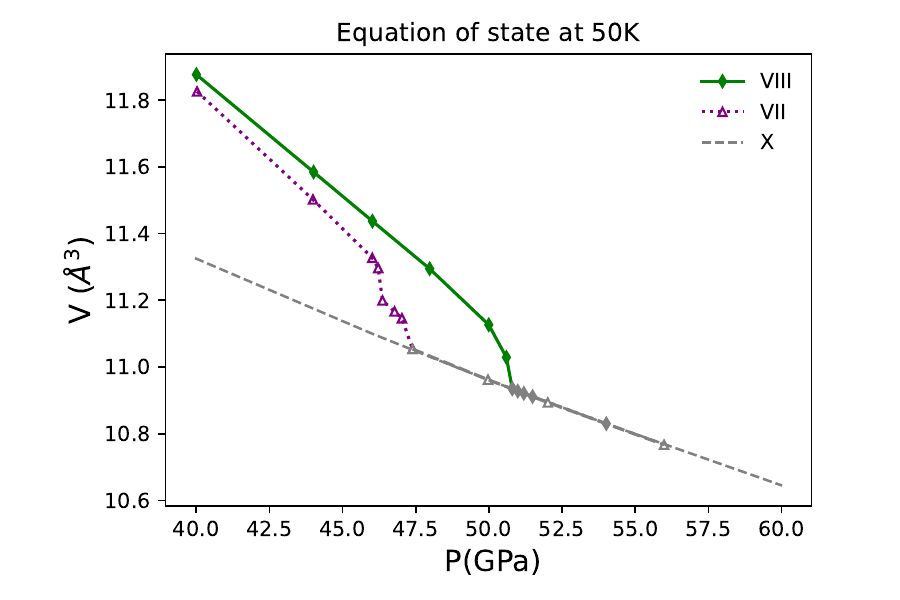}
    \caption{Theoretical high-pressure ice EOS near the critical pressure at 50K shown for ordered phase VIII (green filled diamonds) and disordered phase VII (empty purple upper squares). The volumes are expressed per water molecule. Empty grey triangles and full grey diamonds identify the X phase originated from compressed phases VII and VIII. Grey dashed line refers to phase X minimized with constrained symmetries.}
    \label{fig:EOS_50}
\end{figure}

We present a direct comparison of phases VII and VIII by examining the low-temperature EOS in Fig. \ref{fig:EOS_50}. For ice VII, we use the hydrogen arrangement \emph{struct 0} as a prototypical sample. We notice that, apart from a difference of approximately 3 GPa, the EOS for the two ice phases exhibits similar features. Both phases display a $\simeq 1\%$ volume drop in correspondence with the critical pressure deduced from the order parameter $\Delta$, indicative of a weakly first-order phase transition. At higher pressures, the EOS falls into the X one, by merging exactly with EOS computed by constraining the phase X space group symmetries during the SSCHA minimization. Remarkably, the creation of partially symmetrized hydrogen bonds is also reflected in the EOS of ice VII, which shows the same ragged behavior observed in the order parameter as it approaches the symmetric phase X.

From this analysis, we conclude that the critical pressure of proton symmetrization defining the occurrence of phase X is largely temperature independent. Indeed, both VII-X and VIII-X boundaries do not show any significant temperature dependence, and they are shifted with respect to each other by at most 3 GPa. Therefore, within this pressure range the formation of the symmetric phase X is largely unaffected by static proton disorder, and the SSCHA locates it at $P_c \simeq$ 50 GPa.

\begin{figure}
    \centering
    \includegraphics[width=\columnwidth]{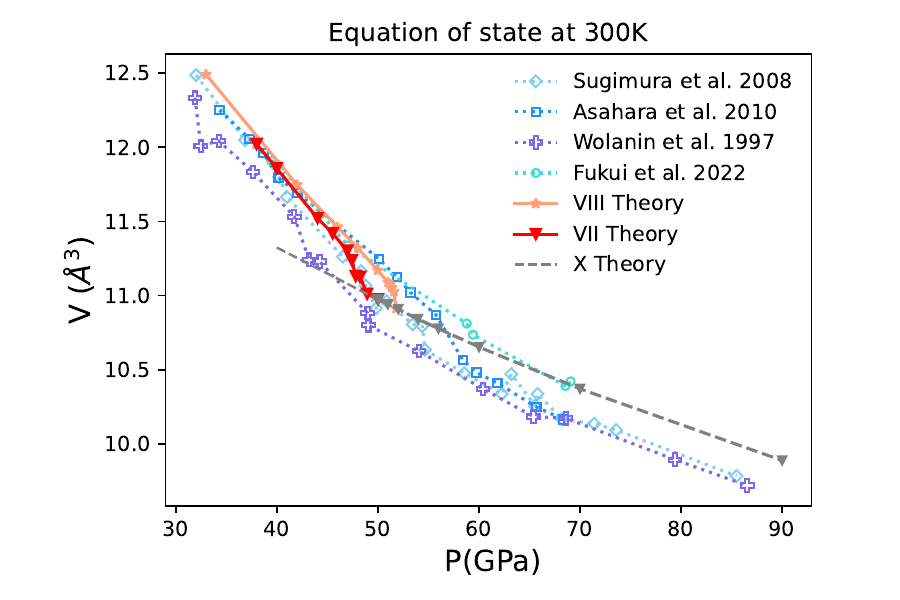}
    \caption{High-pressure ice EOS at room temperature around the critical pressure.
    The theoretical EOSs obtained using the SSCHA for ice VII (full red lower triangles), VIII (full orange stars) are compared with different experimental measurements \cite{Sugimura2008,Wolanin1997,Asahara2010,Fukui2022}. Full grey lower triangles identify the X phase originated from phase VII compression, while the grey dashed line refers to phase X minimized with constrained symmetries.} 
    \label{fig:EOS_300K}
\end{figure}

Experimentally, small variations in the EOS have been observed in the 40-60 GPa pressure range \cite{Wolanin1997,Loubeyre1999,Sugimura2008,Asahara2010,Fukui2022}. However, attributing these features directly to hydrogen bond symmetrization is challenging due to the difficulty in tracking the positions of hydrogen atoms under pressure. To shed light on the experimental findings, we compared our simulated RT EOS around 
the SSCHA critical pressure $P_c$ with some experimental measures in Fig. \ref{fig:EOS_300K}. The temperature independence of the order parameter in Fig. \ref{fig:quantum_SSCHA} suggests that the EOS will show similar features for higher temperatures as well. Indeed, at RT, the system exhibits the same 1$\%$ volume discontinuity at the phase transition observed with SSCHA at 50K. In contrast, the experimental trend appears much smoother, and a direct volume discontinuity is not easily discernible across the transition. Nevertheless, the experimental EOS at T $\simeq$ 300K in Fig. \ref{fig:EOS_300K} matches nicely with the theoretical VII and VIII curves at lower pressure. At the higher pressure side of the transition, our theoretical EOS slightly differs from the experimental determinations, except for the one in Ref. \cite{Fukui2022}. Indeed, the other experimental EOS \cite{Wolanin1997,Sugimura2008,Asahara2010} lead to a P(V) curve shifted with respect to ours, despite showing the same slope, but having less compressible structures.

If we look at the global behavior of our theoretical EOS, we notice a clear change in the slope of the P(V) curve at the transition. The same slope change is apparent also from the experimental data, as one can evince from Fig. \ref{fig:EOS_300K}. However, the non-negligible difference
between the theoretical and experimental EOS is in the pressure location of the cusp. In the experiments, the slope change occurs at around 60 GPa, the same pressure where clear-cut changes also appear in Raman and IR spectroscopies \cite{Aoki1996,Goncharov1996,Song1999}. Instead, in our SSCHA simulations, the transition is located at slightly lower pressures, around 50 GPa. This quantitative difference can be explained by the underlying approximations of the numerical estimate. Those can be attributed to the use of the PBE-DFT functional to derive the interatomic neural network potential.  Additionally, the approximation associated with the use of the SSCHA to deal with quantum nuclear fluctuations could also contribute to the observed differences.

In spite of these differences, based on the SSCHA outcome we can assign the measured EOS to proton-symmetric phases above 60 GPa, since it is outside the discrepancy window and it shows the same slope as the one predicted by our simulations. It follows that if  any intermediate  state differing from phase X or its proton-symmetric variants (\cite{Benoit1998}) exists, its stability domain should be confined to a pressure window below 60 GPa.

Finally, in an effort to characterize the order of the phase transition also with quantum nuclei, we explore the presence of any hysteresis effects by employing the same approach as for the classical case. 

\begin{figure}
    \centering
    \includegraphics[width=0.9\columnwidth]{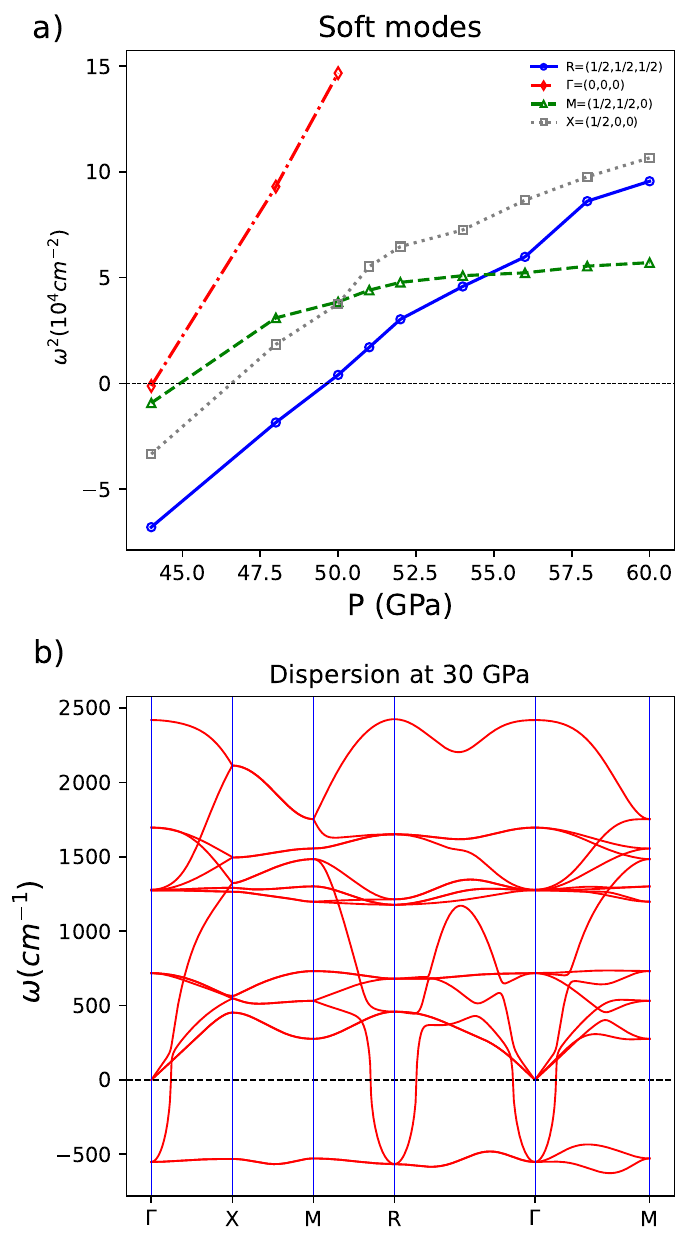}
    \caption{ Panel a): Squared frequency of the soft modes at 300K for the irreducible $\mathbf{q}$-points of the 2x2x2 supercell of phase X as a function of pressure. Phonon frequencies are calculated by diagonalizing the free energy Hessian within the SSCHA framework, including both the third and fourth order terms in the self-energy correction. The phonon instability associated with $\mathbf{q}$=($\frac{1}{2},\frac{1}{2},\frac{1}{2}$) is related to the X-VIII transformation. Panel b): Phonon dispersion along a selected path in the Brillouin zone at 30 GPa and 300K, where ice X is unstable.}
    \label{fig:softmodes}
\end{figure}

In Fig. \ref{fig:softmodes}, we plot the squared frequency of the lowest energy phonons at RT for the four irreducible $\mathbf{q}$-points of the 2x2x2 supercell of phase X computed by SSCHA. Remarkably, at each $\mathbf{q}$-point, we observe a soft mode, whose frequency becomes imaginary at different pressures for different $\mathbf{q}$. During decompression, the first instability encountered corresponds to the soft mode at $\mathbf{q} = (\frac{1}{2},\frac{1}{2},\frac{1}{2})$. Its critical pressure matches the closure point of the VIII-X hysteresis cycle, determined from the order parameter. Indeed, that particular phonon modulates the transformation between the two structures, the unit cell used for ice VIII corresponding to a 2x2x2 supercell of ice X. Nevertheless, with further decompression, all the other $\mathbf{q}$-points show unstable modes and, at a sufficiently low pressure (P $\lesssim$ 44 GPa), an entire phonon branch becomes imaginary (see Fig. \ref{fig:softmodes} (b)). The presence of a fully imaginary dispersionless branch in ice X can be related to the existence of a disordered phase. In fact, disordered structures, with their ideally infinite unit cell, can accommodate all possible configurations associated with phonon instabilities belonging to a dense set of $\mathbf{q}$-points.

The upper and lower pressure edges of the hysteresis cycle for $\ce{H_2O}$ and $\ce{D_2O}$ ices are shown in Fig. \ref{fig:final_plot} (b), defining narrow bands where the VIII-to-X phase transition takes place. We investigate their width as a function of temperature, ranging from T=0K to RT.
For $\ce{H_2O}$ ice, the hysteresis cycle's size exhibits a slight temperature dependence, getting narrower upon heating, decreasing from approximately 2.5 GPa at T=0K to about 1.2 GPa at T=300K. Conversely, for $\ce{D_2O}$ ice, we observed a constant hysteresis width of 1 GPa. In both cases, the relatively small size of the hysteresis cycle suggests a weak first-order phase transition, as found in the classical approximation (see Sec. \ref{sec:classical_results}). Additionally, Fig. \ref{fig:final_plot} compare the SSCHA critical pressures with the experimental data for both isotopes \cite{Pruzan2003,Goncharov1999}. Results from the classical approximation are also included to emphasize the stark contrast between its strong temperature dependence and the flat behaviour observed in the quantum picture.

\section{Conclusions}
\label{sec:conclusion}

In this work, we addressed the long-standing problem of hydrogen bond symmetrization in high-pressure ice performing numerical simulations that combine the SSCHA theory to include NQE with PBE-derived NNP in a temperature range from 0K to RT.

We have shown that a proper description of quantum anharmonicity is crucial to describe the VIII-to-X transition. We found that NQE are so strong to counteract thermal fluctuations and make the critical pressure temperature independent in the range analyzed in our simulations. This result is in agreement with several experimental outcomes \cite{Pruzan1993,Goncharov1996,Song2003} and in striking contrast with the strong non-linear temperature dependence of the critical pressure we observed with classical nuclei.

We observed a small volume drop (about $1\%$) in correspondence with the SSCHA critical pressure of $P_c \simeq$ 50 GPa for the hydrogen bond symmetrization. This is consistent with a weakly first-order phase transition as suggested by the small hysteresis cycle ($\simeq$ 1-3 GPa) observed for all temperatures investigated, independently of isotope substitution.

We have shown that the EOS can be used as a guideline to detect the symmetric phase X. Indeed, both theoretical and experimental EOS show a clear change of slope at the critical pressure. In the experiment, the cusp is located at $\approx$ 60 GPa, in accordance with the spectroscopic signatures for the phase transition. Therefore, the comparison between the theoretical and experimental EOS allows one to exclude the presence of any intermediate state differing from proton-symmetric phases above 60 GPa. Thus, the existence of any intermediate  state with displaced protons should be restricted to a rather narrow window below that value.

The remaining small discrepancy with the experiments in the determination of the critical pressure can be attributed to the underlying approximations used in our calculations, namely the interatomic NNP, the SSCHA treatment of quantum nuclei as well as possible inaccuracies of the DFT approximation.

To check the impact of static proton disorder on the symmetrization pressure, we studied five different realizations of proton disorder in ice VII. 
We have shown that the effect of disorder is negligible, accounting for a 3 GPa critical pressure difference at most. We also reported the existence of a full imaginary branch in ice X at low pressure ($P \leq $ 45 GPa) and RT, already observed in classical simulations \cite{Caracas2008}, suggestive of the tendency of the system to develop disordered local displacements.

This work can be further improved by employing a more exact treatment of quantum anharmonicity as well as a better underlying electronic approximation. Nevertheless, we have shown that advanced quantum theories, like SSCHA, are necessary to unveil not only the structure of high-pressure ice but also its thermodynamic properties, like the EOS. Accessing both structural and thermodynamic quantities plays a pivotal role to interpret the experimental data, and
provides a more comprehensive characterization of the underlying phase transitions.

\begin{acknowledgements}
M.Cherubini acknowledges the CINECA award under the ISCRA initiative, for the availability of high-performance computing resources and support.
L.M. acknowledges the European Union’s Horizon 2020
research and innovation programme under the Marie Skłodowska-Curie grant
agreement 101018714.
\end{acknowledgements}

\section*{Data Availability Statement}
The data that support the findings of this study are available upon reasonable request from the authors.

\appendix

\section{Convergence tests and hysteresis}
\label{app:convergence}

\begin{figure}
    \centering
    \includegraphics[width=0.9\columnwidth]{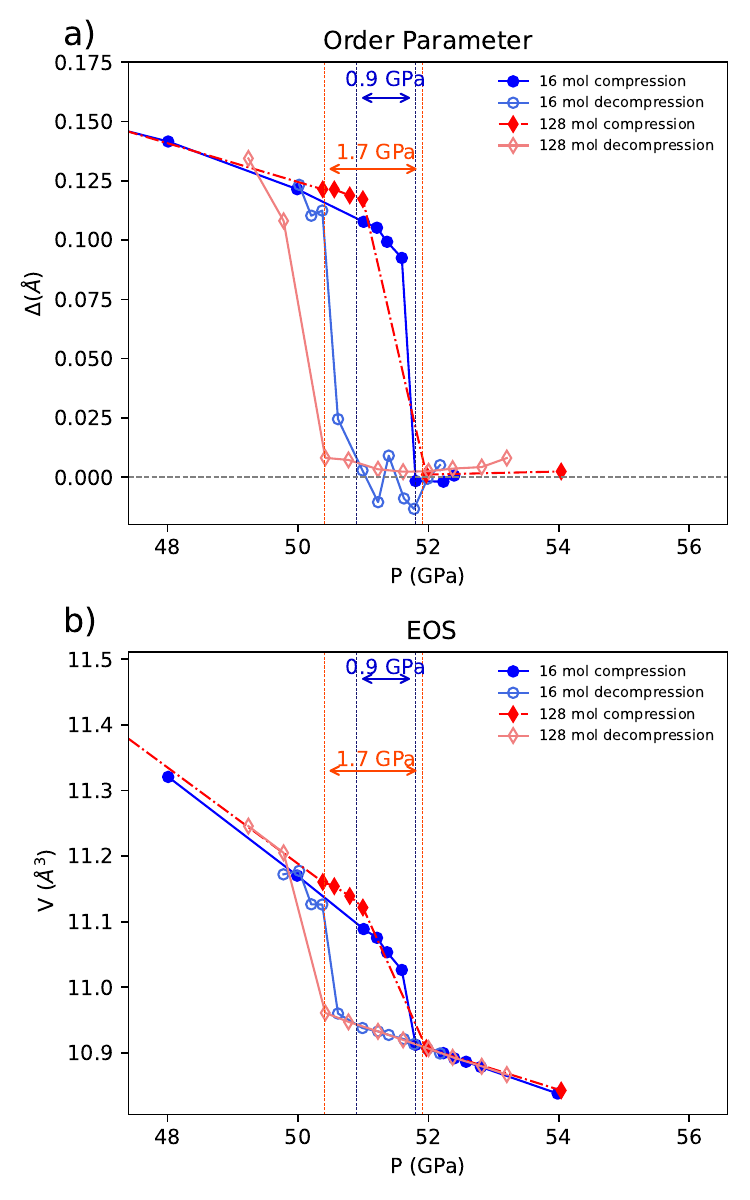}
    \caption{Hysteresis cycle for ice VIII at 300K for different supercell sizes. Panel a): Order parameter $\Delta$ as a function of pressure for the unit cell with 16 water molecules (blue circles) and its 2x2x2 supercell (red diamonds). Compression and decompression runs are represented by filled and empty points, respectively. Dashed vertical lines indicate the hysteresis cycle, whose size is written above the double-sided arrows. Panel b): Equation of state computed in the same conditions as in panel a). Colors and symbols retain the same meaning.}
    \label{appfig:Hysteresis_VIII_300}
\end{figure}

In the main text, we presented simulations of water ice using 16 $\ce{H_2O}$ ($\ce{D_2O}$) molecules. This is the same number of molecules adopted in pioneering PIMD simulations focused on studying H-bond symmetrization \cite{Benoit1998,Benoit2002}. We checked the convergence of the order parameter $\Delta$ and the equation of state, crucial properties for identifying the VII/VIII-X phase transition. We compared the results obtained in the unit cell with 16 water molecules to those in a 2 $\times$ 2 $\times$ 2 supercell containing 128 water molecules. We analyze two temperatures, T=50K and T=300K, studying also the hysteresis cycle width as a function of the cell size.

In Fig. \ref{appfig:Hysteresis_VIII_300}, we show the pressure dependence of the order parameter and the EOS, both in the compression and in the decompression runs at RT for ice VIII. The presence of hysteresis, independent of the number of molecules in the simulations, dismisses the possibility of its origin being attributed to size effects. Although the hysteresis cycle width exhibits a slight dependence on the cell size, this difference is tiny (less than 1 GPa) compared to the pressure scale where hydrogen bond symmetrization occurs. The predicted critical pressures, derived by means of the order parameter or from the EOS, remains nearly identical in both simulation cells during compression runs. Moreover, also the absolute values of the order parameter and the P(V) curve show no significant dependence on the cell size.

\begin{figure}
    \centering
    \includegraphics[width=0.9\columnwidth]{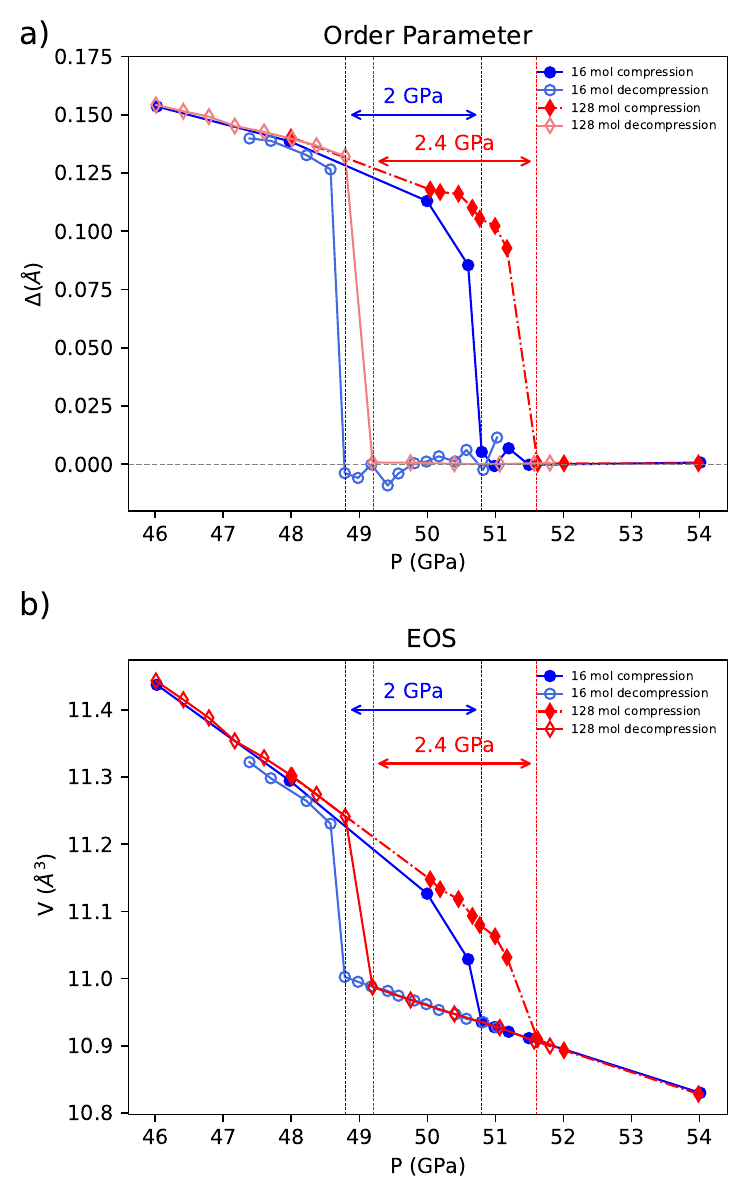}
    \caption{Critical pressure of ice VIII at 50K as a function of the supercell size. Panel a): Order parameter $\Delta$ as a function of pressure for the unit cell with 16 water molecules (blue circles) and its 2x2x2 supercell (red diamonds). Compression and decompression runs are represented by filled and empty points, respectively.  Dashed vertical lines highlight the width of the hysteresis cycle. Panel b): Equation of state computed in the same condition as in panel a). Colors and symbols maintain the same meaning.}
    \label{appfig:Hysteresis_VIII_50}
\end{figure}

We carried out a similar analysis at T=50K (Fig. \ref{appfig:Hysteresis_VIII_50}). Compared to RT, we observe a slightly larger difference in the results for the 16 and 128 water molecules simulations. By enlarging the cell size, there is a small increase in the absolute value of the critical pressure ( $\simeq$ 0.7 GPa). However, except in the discrepancy region, both the EOS and the order parameter remain almost identical as a function of the cell size. As for the RT simulations, the hysteresis cycle size shows a negligible dependence on the supercell dimension ($\simeq$ 0.4 GPa). These differences are tiny compared to the absolute value of the transition pressure. Thus, we consider the 16 water molecules unit cell to adequately satisfy the convergence requirements.

\begin{figure}
    \centering
    \includegraphics[width=0.9\columnwidth]{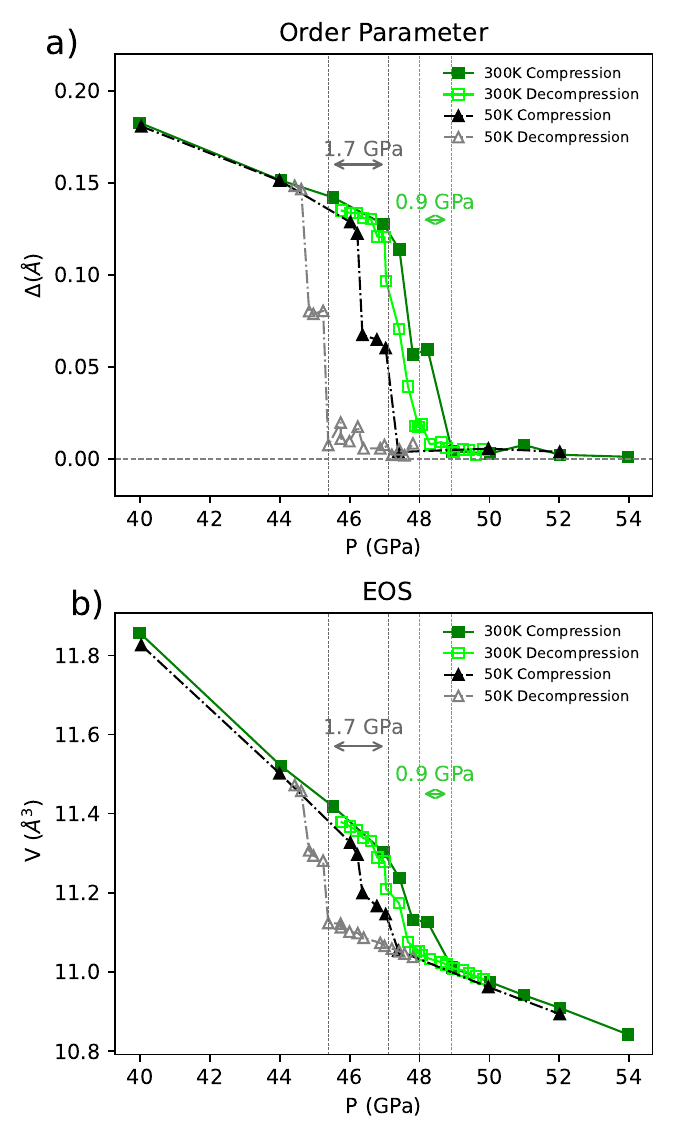}
    \caption{Hysteresis cycle for ice VII at 50K and 300K in the unit cell with 16 water molecules. In all the panels, filled and empty symbols indicate for the compression and the decompression runs, respectively. Panel a): Global order parameter $\Delta$ as a function of pressure at T=50K (dark upper triangles) and room temperature (green squares). The hysteresis cycle is shown by the vertical dashed lines. Panel b): Equation of state computed in the same condition as in panel a). Colors and symbols retain the same meaning.}
    \label{appfig:Hysteresis_VII}
\end{figure}

We then turn our attention to exploring the presence of hysteresis effects in ice VII. In Fig. \ref{appfig:Hysteresis_VII}, we show the global order parameter $\Delta$ and the EOS at T=50K and RT for the proton-disordered realization labeled \emph{struct 0} in Fig. \ref{fig:different_VIIs}. 
We recall that for the disordered structures, the global order parameter is obtained by averaging the local ones $\Delta_i$ associated with each hydrogen atom in the structure (see Sec. \ref{sec:quantum_results}).
For ice VII, the hysteresis cycle width increases for lower temperatures, ranging from $\simeq$ 0.9 GPa at 300K to $\simeq$ 1.7 GPa at 50K, mirroring the trend observed in ordered ice.

\begin{figure}
    \centering
    \includegraphics[width=0.9\columnwidth]{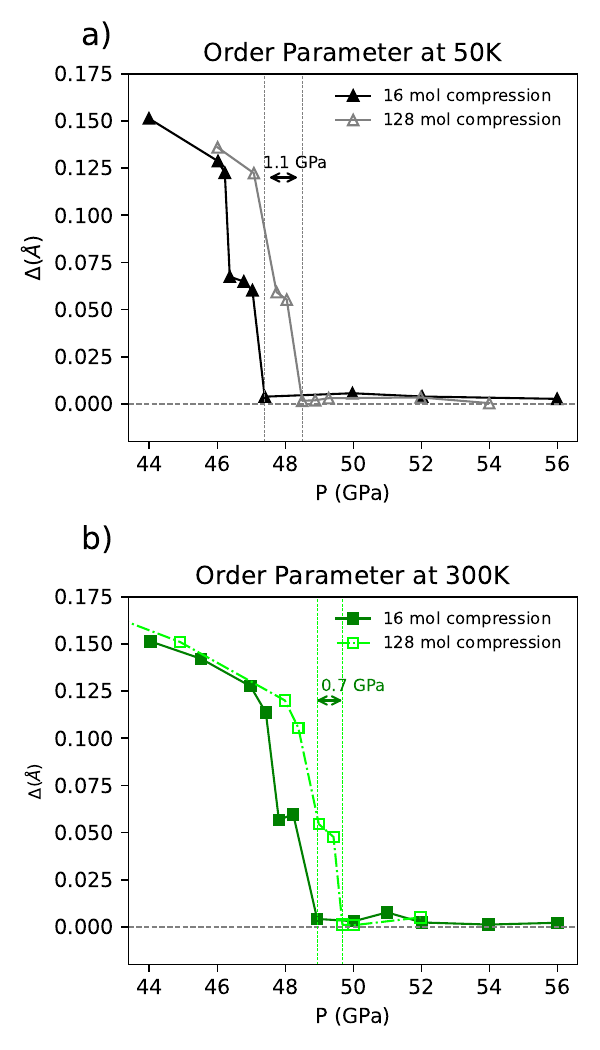}
    \caption{Convergence test for ice VII at 50K (panel a)) and 300K (panel b)). Filled and empty symbols show the global order parameter $\Delta$ for the unit cell with 16 water molecules and the 2x2x2 supercell with 128 molecules, respectively. The vertical dashed lines indicate the corresponding critical pressure, whose difference between unit cell and supercell is written above the double-shaped arrows.}
    \label{fig:Convergence_VII}
\end{figure}

Finally, we study the dependence of the proton symmetrization pressure as a function of the cell size in phase VII. It is important to underline that we are simply using a 128-molecule 2x2x2 supercell of our unit cell, and we are not considering a realization of disorder in a cell with 128 water molecules. By increasing the system size, the hydrogen bond symmetrization pressure increases (see Fig. \ref{fig:Convergence_VII}), as for the ordered ice VIII. Nevertheless, the shifts of 1.1 GPa and 0.7 GPa, at 50K and 300K, respectively, are relatively small compared to the pressure scales under investigation. Therefore, even for the disordered phase, we achieve the required accuracy and convergence by employing the unit cell with only 16 water molecules.

\section{The free energy Hessian}
\label{app:hessian}

The SSCHA dynamical matrix, obtained through free energy minimization, is not suitable for describing real phonons because it is positive definite by construction \cite{Monacelli2021}. This property means that it cannot capture phonon instabilities. However, physical phonons can be obtained by computing the free energy curvature in the centroid position. 
Since we need phonons only to identify instabilities in phase X, it is sufficient to use the static approximation of the self-energy.

\begin{figure}
    \centering
    \includegraphics[width=0.9\columnwidth]{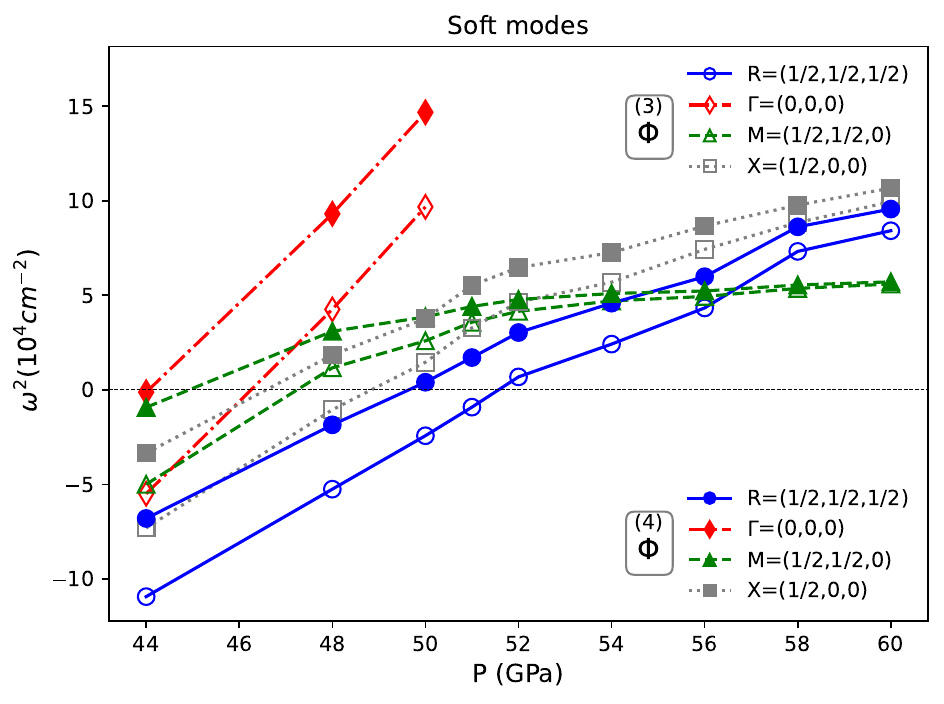}
    \caption{Role of four-phonon scattering in ice X at 300K. The simulations were carried out by employing a 2x2x2 supercell of phase X containing 16 water molecules. All the soft modes in the irreducible $\mathbf{q}$-points of the supercell are plotted. Full symbols represent frequencies computed with the full free energy Hessian in the static approximation, while empty symbols indicate frequencies computed with only the third-order term in the self-energy expression \cite{Monacelli2021}.}
    \label{fig:Hessian}
\end{figure}

\begin{figure}
    \centering
    \includegraphics[width=\columnwidth]{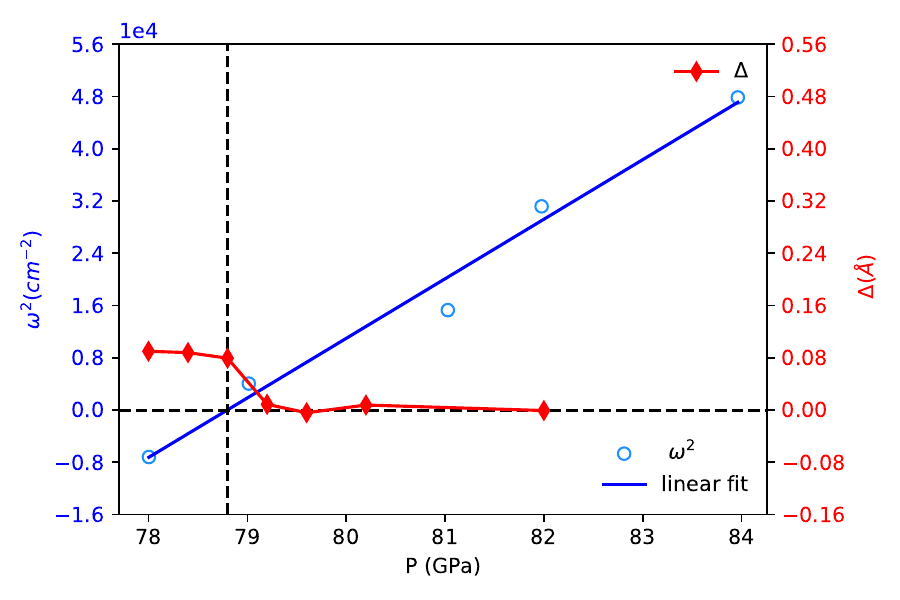}
    \caption{Comparison between different methodologies to compute the closure point of the hysteresis cycles in the classical approximation at RT. Filled dots represent the squared frequency for the lowest energy phonon at $\mathbf{q} = (\frac{1}{2},\frac{1}{2},\frac{1}{2})$. The blue line is a linear fit used to identify the pressure where the phonon becomes imaginary, indicated by the vertical dark dotted line. Red filled diamonds show the order parameter $\Delta$ in decompression. Left blue and right red axis refer to the squared frequency and the order parameter, respectively.}
    \label{appfig:hessian_hysteresis}
\end{figure}

In Fig. \ref{fig:Hessian}, we show the comparison of phonon frequencies as a function of pressure computed by including or excluding the fourth-order force constant matrix in the self-energy expansion (see Ref. \cite{Monacelli2021} for further details) using 50000 configurations. The inclusion of all terms to calculate the free energy curvature is crucial for obtaining a good description of phonon instabilities. Neglecting the fourth-order term reduces the stability domain of ice X by more than 1 GPa.

In Sec. \ref{sec:classical_results}, we highlighted that we opted to use the appearance of imaginary modes to locate the closure point of the hysteresis cycle instead of relying on the order parameter for computational efficiency reasons. In Fig. \ref{appfig:hessian_hysteresis}, we show that both methods yield consistent results for ice VIII at RT with classical nuclei. We fitted the squared frequency of the lowest energetic phonon at $ \mathbf{q} = (\frac{1}{2},\frac{1}{2},\frac{1}{2})$, associated with the VIII-X instability, as a function of pressure. The pressure at which it vanishes matches perfectly with the pressure where the hydrogen atoms leave the O-O midpoint in the decompression run. Similar checks have been performed for the other cases presented in the main text.

\begin{figure}
\centering\includegraphics[width=\columnwidth]{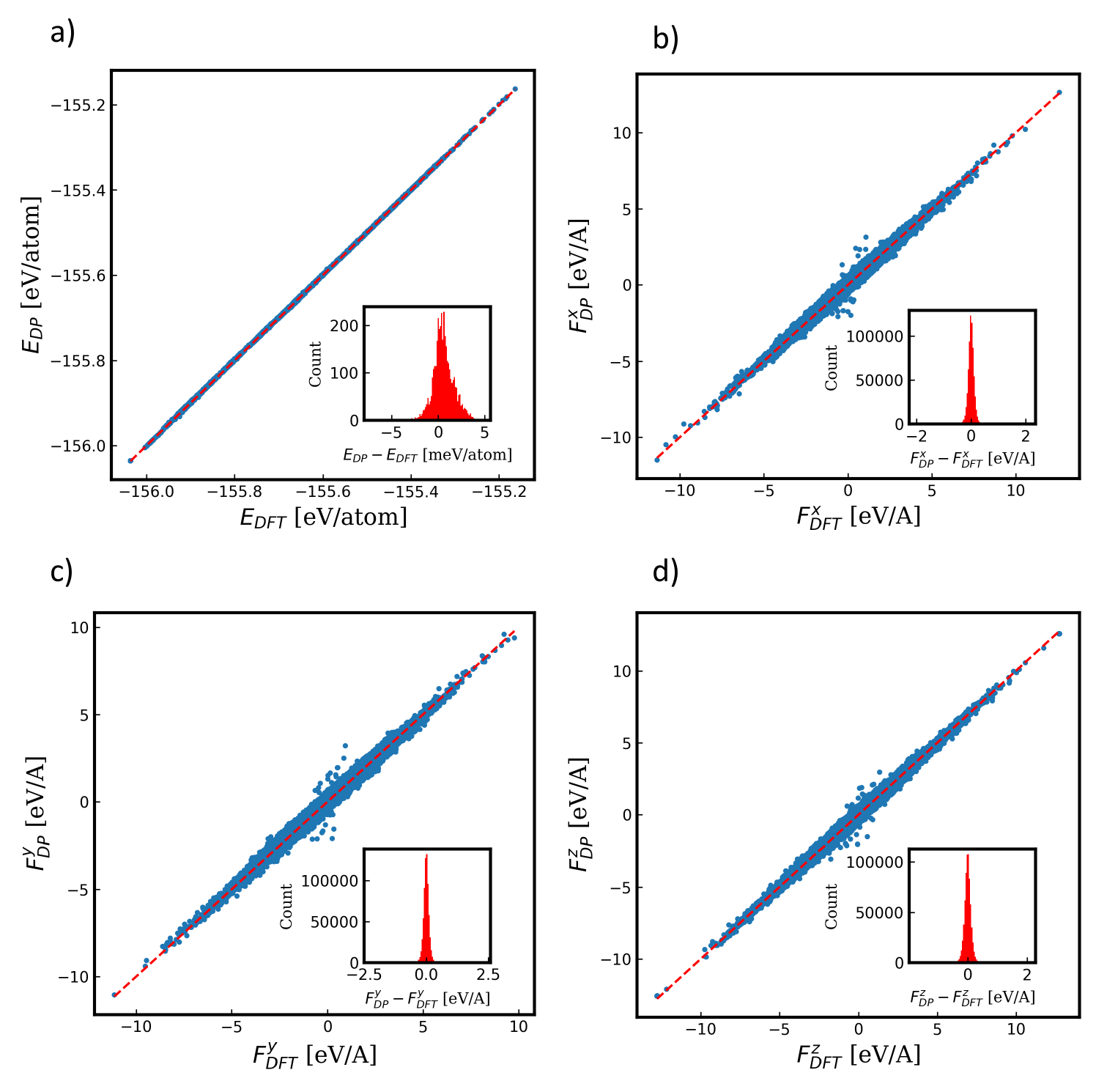}
    \caption{ (a) Comparison of energies predicted using DP model ($E_{\mathrm{DP}}$) with the reference PBE-DFT results ($E_{\mathrm{DFT}}$) on the training dataset. The inset shows the error distribution. (b-d) Comparison of the Cartesian components of force predicted using DP model ($F^{x/y/z}_{\mathrm{DP}}$) with the reference PBE-DFT results ($F^{x/y/z}_{\mathrm{DFT}}$) on the training dataset. The insets show the error distributions.
     }
    \label{fig:dp}
\end{figure}

\section{The neural network potential}
\label{app:DP}
The neural network potential used in this work is a short-range Deep Potential~\cite{zhang2018deep} model trained on DFT data. The cutoff radius for short-range interaction is 6\AA.  The DFT dataset is collected by the active learning agent DPGEN~\cite{zhang2020dp} in temperature range $[50, 330]$K and pressure range $[20, 120] $GPa.  Supercell of different sizes are also used for better generalizability. The number of atoms in the supercell varies from 48 (16 $\ce{H_2O}$  molecules) to 384 (128 $\ce{H_2O}$  molecules). 
The final dataset contains 4022 atomic configurations, where 298 configurations are collected from path-integral molecular dynamics trajectories. The others are collected from classical molecular dynamics trajectories. 

Fig.~\ref{fig:dp} shows the performance of our model on the dataset.  Typical prediction error in potential energy is smaller than 5meV/atom. The standard deviation of the error in energy is 1meV/atom. Meanwhile, typical prediction error in the Cartesian components of force is smaller than 1eV/$\text{\AA}$. The standard deviation of the error in force components is 0.1eV/$\text{\AA}$.

\bibliography{main}

\end{document}